\documentclass{aipproc}
\def\selectedlayoutstyle{8x11single}
\layoutstyle\selectedlayoutstyle
\SetInternalRegister\hbadness{8000} % pseudo latin isn't breaking very well :-)
\newcommand\doingARLO[2][]{%
  \ifx\mmref\undefined #1\else #2\fi
}
\newcommand{\degr}{^{\circ}}

\begin{document}

\title[Magnetic fields in our Galaxy]
      {Magnetic fields in our Galaxy: How much do we know?\\
	(II) Halo fields and the global field structure}


\author{JinLin Han}{
address={The partner group of MPIfR},
address={National Astronomical Observatories, Chinese Academy of
Sciences}, 
address={Jia-20 DaTun Road, ChaoYang District, Beijing 100012, China},
address={email: hjl@bao.ac.cn}
}

% \copyrightholder{Acoustical Scociety of America}
\copyrightyear  {2001}

\begin{abstract}
I review the large scale global magnetic field structure of our Galaxy,
using all information available for disk fields, halo fields and
magnetic fields near the Galactic center (GC). Most of the knowledge was
obtained from the rotation measures (RMs) of pulsars and extragalactic
radio sources. In the local disk of our Galaxy,  RM and dispersion
measure (DM) data of nearby pulsars yield the strength of regular field
as 1.8$\mu$G, with a pitch angle of about $8\degr$, and a bisymmetric spiral
structure. There are at least four, maybe five, field reversals
from the Norma arm to the outskirts of our Galaxy. The regular fields
are probably stronger in the interarm regions. The directions of regular
magnetic fields are coherent along spiral arms over more than 10kpc.
The regular fields get stronger with decreasing radius from the GC.
In the thick disk or Galactic halo, large scale toroidal magnetic
fields, with opposite field directions in the Southern and Northern
Galaxy, have been revealed by the antisymmetric RM sky towards the inner
Galaxy. This signature of the A0 dynamo-mode field structure is
strengthened by the indication of a poloidal field of dipole form,
that is the transition of the RM signs probably shifted from $l\sim0\degr$
to $l\sim+10\degr$. The local vertical field is probably a part of
this dipole field. The field structure of the A0 dynamo-mode strikingly
continues towards the region near the GC. We predict that the direction
of the dipole fields near the GC should point towards the Southern pole.
In short, the magnetic fields in the Galactic disk have a bisymmetric
spiral structure of primordial nature, while in the halo and near
the GC the A0 dynamo seems to dominate, so that the fields consist
of toroidal fields with opposite directions below and above the
Galactic plane and poloidal fields of dipole form.
\end{abstract}

\date{\today}

\maketitle

\section{Introduction}
The magnetic field structure in the Milky Way Galaxy is not yet fully known.
It probably will never be observed completely, analogous to the situation
of mapping the large scale structure of the universe. However, our Galaxy
is a very special case for study, partly because we can see more
``trees'' though not the ``forest''. One can observe many more details
of the magnetic fields, and study their roles in star formation regions
\cite{ree87}.  The details of local field structure and strength are
very fundamental to understanding cosmic rays \cite{bc00}, especially
those of high energy. The diverse polarized features observed in the
diffuse radio emission \cite{gdm+01, gld+99, ufr+98, dhjs97} are closely
related to the small- and large-scale of magnetic fields. The fields
are also important to the hydrostatic balance \cite{bc90} and stability.
While in the Galactic disk, the spiral structure is not yet clear,
measuring the regular magnetic field structure could be an approach
to mapping the Galaxy.

The major advantages of studying the magnetic field of our Galaxy are
the fact that the Galaxy fills the sky and that a very number of pulsars
and polarized extragalactic radio sources can be used as probes of
the three dimensional magnetic field structure. These
are unique to our Galaxy and extremely powerful for the halo field study.
For the second largest galaxy in the sky, M31, there are only 21
bright polarized background radio sources\cite{hbb98}.

Here I review the new progress obtained mainly in the last decade,
and point to several earlier reviews (e.g. Sect.3 of \cite{wk93},
\cite{sfw86}) which clearly show the situation about 10 years ago.
This is a companion paper for the review of observational facts of
disk fields within a few kpc from the Sun \cite{han01}. This paper
will focus on the field structure in the halo and near the Galactic
center, with brief discussion of the disk fields so that the global
field structure can be delineated.

\section{Magnetic fields in the Galactic disk}

Significant progress has been made in the last decade on the magnetic
fields in the Galactic disk, mainly because of many pulsars newly
discovered in the nearby half of the whole Galactic disk
\cite{mld+96, lml+98, mlc+01} and extensive observations of pulsar
RMs \cite{hl87, rl94, hmq99}. Analyses of pulsar RM data
\cite{hq94, rl94, id98, hmq99, hmlq01} yield most definite key
parameters of the regular magnetic field, namely the pitch angle of
local fields, the field strength, and the field reversals.

Determining the local magnetic field was the main objective of measuring the
polarization of star light. Using the largest data set of 7500 stars
\cite{am89, hei96a}, local magnetic fields were found to be
concentrated in the spiral arms and directed along the axes of the
arms, with a pitch angle of $p=-7.2\degr \pm 4.1\degr$. About a decade
ago, a few dozen pulsar RMs were not enough to determine the field
pitch angle more accurately than 10$\degr$. After Hamilton \& Lyne
obtained the RMs of 185 pulsars\cite{hl87}, Han \& Qiao obtained
the pitch angle $p=-8.2\degr\pm0.5\degr$ from model-fitting to
the data of carefully selected pulsars within 3 kpc\cite{hq94}. The
result was confirmed later using more pulsar RM data \cite{id98, hmq99},
and is consistent with the value from optical data. Nowadays, there is
no longer any doubt on the spiral feature of the local regular field,
which has a pitch angle, $p = -8\degr$, with a maximum uncertainty
of $2\degr$.

Talking about the field strength, we have to be aware of the difference
between the total (rms) field strength, the strength of the regular field,
the average field strength over a line of sight, and the maximum
strength of the reversed field model. Total fields obviously are
stronger in the arm regions, mainly contributed by random
fields. That is about 5 $\mu$G in the vicinity of the Sun. However,
regular fields are stronger in the interarm regions \cite{hq94, id98}.
The average field strengths {\it directly} determined from pulsar DM
and RM are mostly in a range of $1\sim2\mu$G, with a maximum about
5$\mu$G (see Fig.1), suggesting a regular field of $1\sim2\mu$G
and a random field about $5 \mu$G. These are exactly the values
obtained by sophisticated analyses \cite{os93, rk89, hq94}. Note that
the magnetic field energy stored in the random component is 3.7 times than
the regular field \cite{hq94}, indicating that the random field
always dominates. Our new measurements of magnetic field in the
Norma arm\cite{hmlq01} show that the regular fields themselves could
be as strong as $5 \mu$G, indicating that the field strength probably
increases smoothly towards the Galactic center.
\begin{figure}[h]
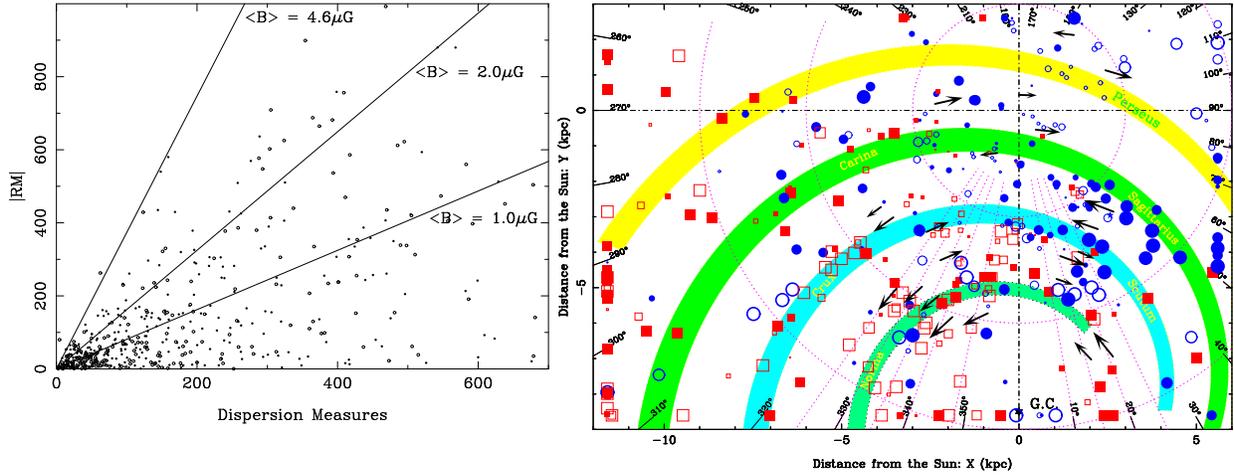

\resizebox{7.2cm}{!}{\includegraphics[angle=270]{han_rm_dm.ps}}
\resizebox{9cm}{!}{\includegraphics[angle=270]{han_rm_xy.ps}}
\caption{{\it Left:} Pulsar rotation measure values plotted against
 pulsar dispersion measure values, directly show the
field strength averaged over the paths from pulsars to us. \hspace{1mm}
{\it Right:} The distribution of pulsar rotation measures projected
onto the Galactic plane. Square symbols represent new measurements.
Open symbols represent negative RM values.  The field directions along
the four spiral arms are indicated by arrows.}
\end{figure}

Pulsar rotation measures can tell the directions of the averaged
field, in contrast to the orientation obtained from polarization of
starlight or mapping at radio bands. The local regular magnetic
field is pointing towards $l\sim82\degr$. 
There are at least four, maybe five, field reversals (see Fig.1)
from the Norma arm to the outskirts of our Galaxy\cite{hmlq01, hmq99,
 id98, rl94, hq94}. By comparison of the RM values of distant pulsars
with those of extragalactic radio sources, at least one, probably
two field reversals have been identified near and beyond the Perseus arm
\cite{ls89, ccs+92, hmq99}. In the inner Galaxy, the fields
first reverse their direction at about 0.2 kpc inside the solar
ring, near the Carina-Sagittarius arm. The fields are reversed back
again near or interior to the Crux-Scutum arm, as first shown by
pulsar RM data \cite{rl94}. New RM observations of more distant
pulsars suggested that the fields near the Carina-Sagittarius arm
and the Crux-Scutum arm are coherent in direction over more
than 10 kpc along the spiral arms. From the most recent data we have
identified the coherent counterclockwise fields near the Norma arm,
indicating a third field reversal \cite{hmlq01, hmq99}.

Compared to that of external galaxies, the magnetic fields in the
disk of our Galaxy is exotic because of these field reversals.
Similar phenomena have not been observed in external galaxies \cite{bec01}.
A proper model to available data is the only way when astronomers
have only the parts but like to understand the entire story.
Three models were proposed for the global structure of magnetic
fields in the disc of our Galaxy: the concentric ring model
\cite{rk89, rl94}, the axisymmetric spiral (ASS) model \cite{val96}
and a bisymmetric spiral (BSS) \cite{sk80, sf83, hq94, id98}.
\begin{figure}[h]
  \includegraphics[height=40mm]{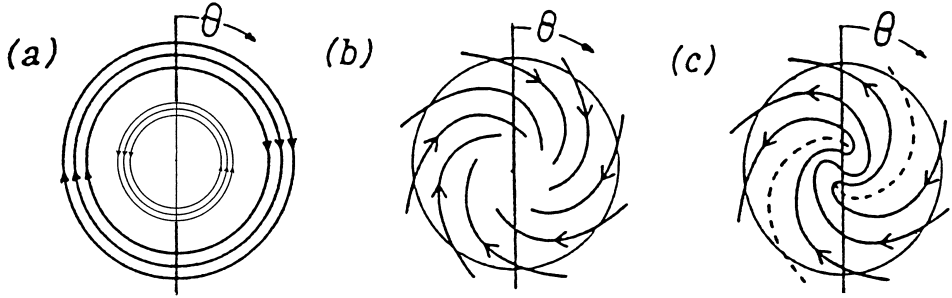}
  \caption{A sketch of three models for Galactic magnetic fields
(modified from \cite{fs90}), namely, (a) the concentric ring model,
(b) the axi-symmetric spiral model, and (c) the bi-symmetric spiral
model.}
\end{figure}

Only the concentric-ring model \cite{rk89, rl94} has a pitch angle
of zero  but the observed pitch angle of fields (see above) argues
for a spiral form of the fields. Nevertheless, this model allows
the fields to be reversed in different ranges of Galacto-radius,
so this model explain the RM data for the zero-order of approximation.
In the ASS model \cite{val96}, the reversed field occurs only in
the range of Galacto radius from 5 to 8 kpc. No field reversals are
allowed beyond 8 kpc from the GC center or within 5 kpc of the GC.
Field reversals beyond the solar circle or interior to the Crux-Scutum
arm are difficult to reconcile with this model. The spiral nature
of the regular field and the field reversals strongly support the
BSS model for the disk fields in our Galaxy. Model fitting to
available pulsar RM data \cite{hq94, id98} have confirmed this
conclusion.

Although only the BSS model has survived confrontation with RM data
and star polarization data, it may be worth reminding that, though
the model fits the available data, it could fail to fit data from 
these part of the disk as yet unexplored. There are two possibilities
for farther tests of the model. One is to check closely the coherence
of field directions; the other is to determine the fields in more
distant regions. The essential step for both avenues is to obtain
more pulsar RM data in the distant parts of the galactic disk.

\section{Magnetic fields in the Galactic halo}

As was shown by previous radio background observations, our Galaxy consists
of two components\cite{bkb85}, a thin disk and a thick disk. The magnetic
field in the thin disk is more dominated by the spiral structure, as we
discussed above. The thick disk has a scale height of 1.5 kpc near
the solar circle \cite{sk80, hq94}. We refer to the thick disk, together
with the more extended component, as the Galactic halo. The magnetic fields
in the thin disk diffuse into the halo, as shown by the spurs or
plumes emerging from the Galactic plane \cite{hssw82, dhr+98} and
as needed by dynamo actions.

Our Galaxy is the largest visual galaxy
in the sky, and the RMs of extragalactic radio sources (EGRS) and pulsars
are the best probes for the halo field. Our Galaxy provides a unique
opportunity to study halo field structure, on either small or large
scales.
However, the RM sky has many features \cite{sk80} related to
local high-latitude phenomena, such as, the North Polar Spur region
(Loop I), Region A (loop II), both of which are probably superbubbles
or supernova remnants. Faint large-scale $H_{\alpha}$
filaments sometimes also affect the RM distribution \cite{hmq99}.
Halo fields have not been well studied well, mainly because of these
dominant local features.

After carefully filtering out deviant points, we found the antisymmetry
of the rotation measure sky in the {\it inner} Galaxy\cite{hmbb97}. Such a
highly symmetric pattern could not be produced by just the coincidence
of many local perturbations, instead, it is of large, probably galactic,
scale. Such an antisymmetric sky suggests a magnetic field in
the halo with opposite directions in the northern and southern Galactic
hemispheres. We noticed later that Andreasyan \& Makarov independently
made a similar suggestion\cite{am88}. A very important fact is
that such an antisymmetry only occurs at high latitudes for the 
inner part of the Galaxy, indicating that it is related to the field
structure mainly in the Galactic halo. The field structure so revealed
is amazingly consistent with that produced by an A0 dynamo.
Together with the vertical field near the Galactic center (see below),
we believe that an A0-mode dynamo is operating in the Galactic halo.
This is the first time that a dynamo mode has been identified on a
galactic scale.

\begin{figure}[h]
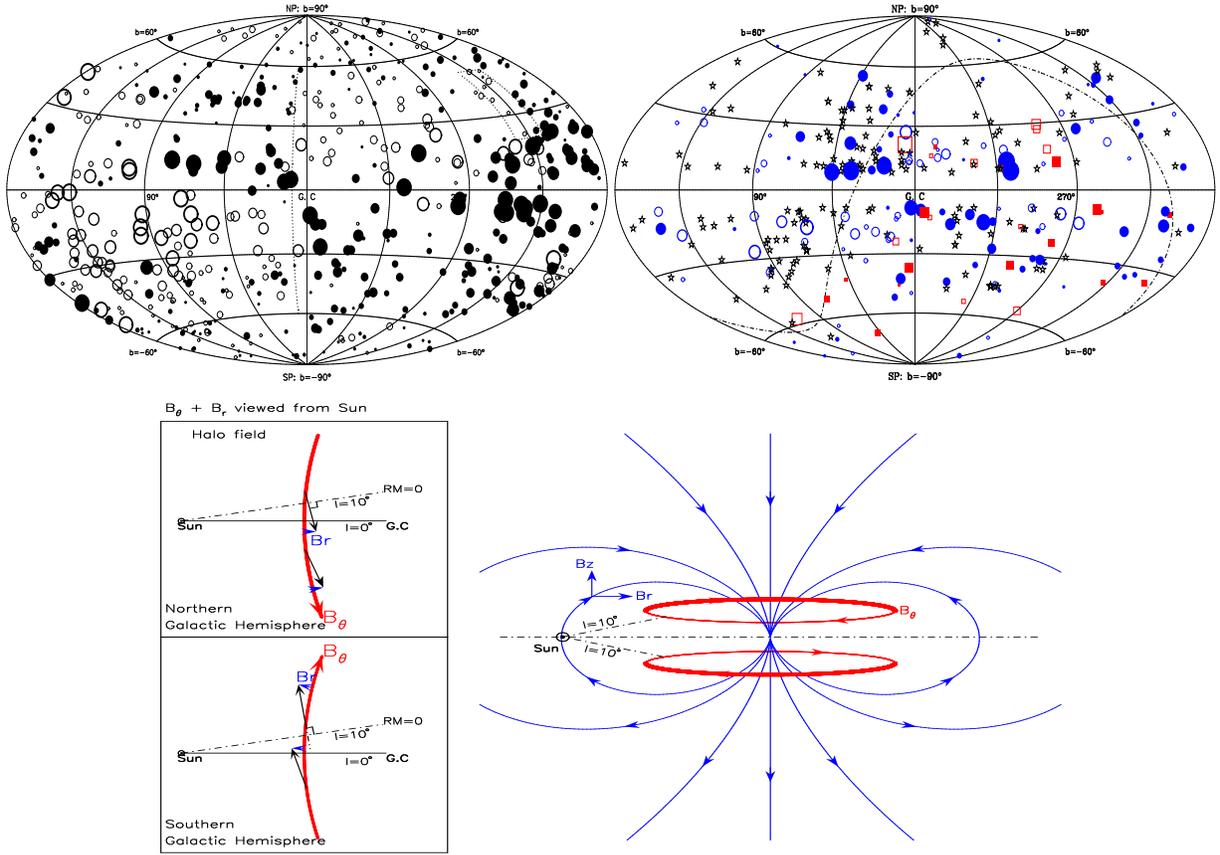
		% 	Fig.1
\begin{tabular}{c}
\mbox{\includegraphics[height=8cm,width=5cm,angle=270]{han_ers.ps}}
\mbox{\includegraphics[height=8cm,width=5cm,angle=270]{han_psr_lb.ps}}
 \\
\mbox{\includegraphics[height=12cm,width=6cm,angle=270]{han_B_model.ps}}
\end{tabular}
\caption{The antisymmetric rotation measure sky of extragalactic radio
sources ({\it upper left}) and pulsars ({\it upper right}) directly indicates
the reversed direction of the azimuthal magnetic fields in the southern
and northern hemisphere of our Galaxy. The filled symbols represent
positive RMs, indicating the averaged field towards us. The effect of
the dipole field on the longitude transition of RM signs ({\it lower left})
and the magnetic field configuration of an A0-mode dynamo ({\it lower right})
are shown for comparison.}
\end{figure}

Two facts further support the argument for the dynamo origin of the
antisymmetric rotation measures. One is the local vertical fields.
After all local perturbations were discounted, we found that the
vertical fields have a strength of 0.2$\sim$0.3 $\mu$G and point
from the South Galactic pole to the North \cite{hq94, hmq99}. If
this is a part of the dipole field expected from the A0 mode dynamo,
then the longitude transition of RM signs should be slightly shifted
from $l\sim0\degr$ to $l\sim$ a few degrees, depending to the strengths
of the dipole fields and the toroidal fields. The halo toroidal fields
have a strength of about 1 $\mu$G \cite{hmq99}. The transition shift,
though marginally shown by the available RM data of extragalactic radio
sources and pulsars, was first noticed  by Han et al. in 1999 \cite{hmq99}.
We understand that this second fact needs more data to confirm, which
we are currently working on.

In short, the RM sky demonstrates an A0 dynamo acting in the halo,
producing azimuthal fields with opposite directions in the
southern and northern galactic hemispheres and also dipole
fields. All pieces of evidence are nicely consistent for these
field components of the A0-mode dynamo field structure. Computer
simulation of the toroidal and dipole fields, together with the
electron distribution model\cite{tc93} have confirmed all these
features.

\section{Magnetic fields near the Galactic center}

Two aspects of the magnetic fields near the Galactic center should
be discussed -- first how the disk fields continue towards the
center, and second, the possible presence of poloidal fields and
toroidal fields there and any possible connections with the fields
in the Galactic halo. On the continuation of disk fields,
there is a knowledge gap to fill. We have only recently determined
the field structure near the Norma arm and have no idea
at all about the fields interior to this arm. It is not clear whether
the strong magnetic fields observed in the central molecular
zone \cite{ncr+01, ndd+00} are continued and connected to the
large scale fields in the disk. It is the case for the nearby galaxy
NGC 2997\cite{hbe+99}.

\begin{figure}[t]
\resizebox{71mm}{!}{\includegraphics{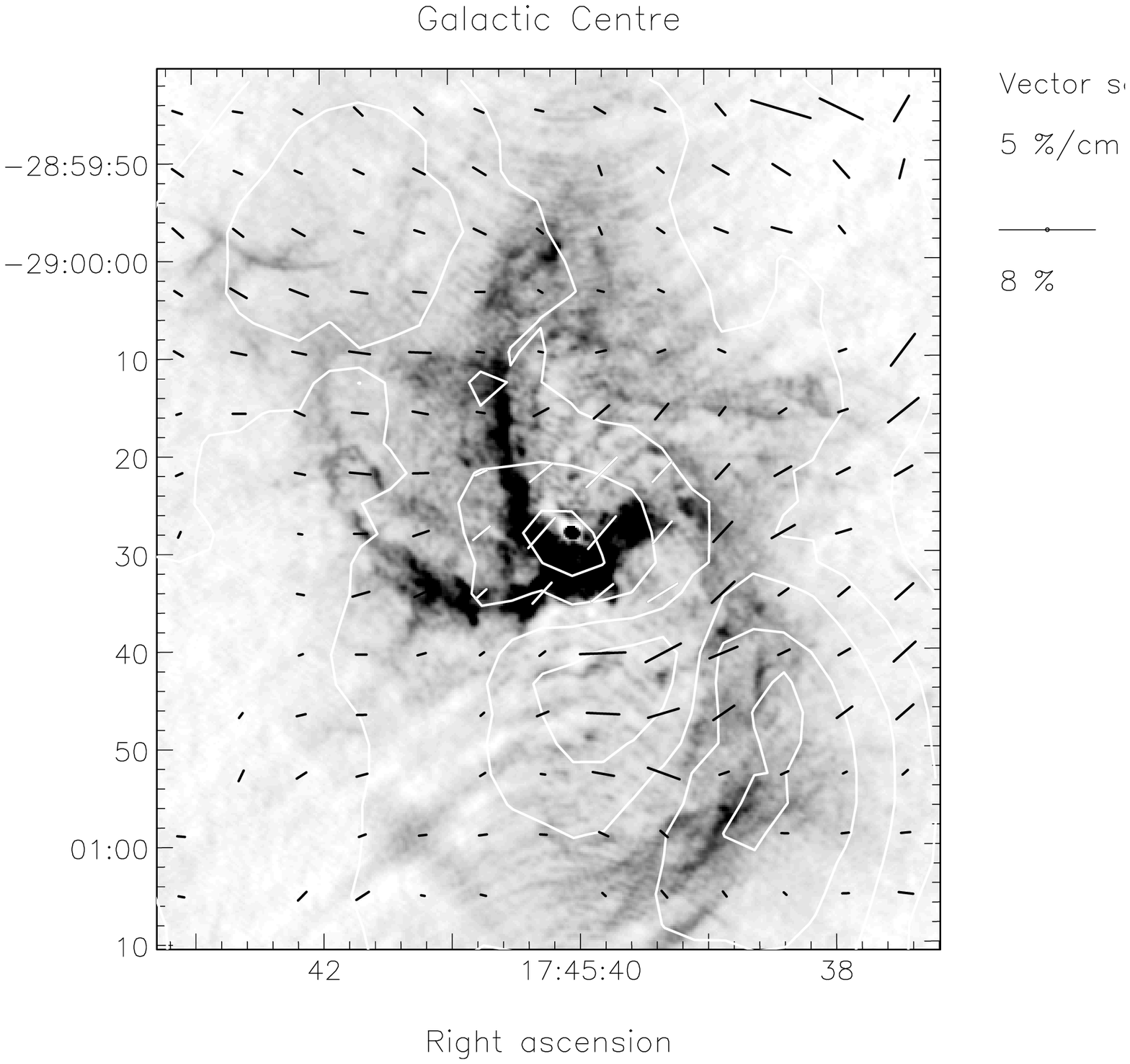}} \hspace{2mm}
\resizebox{74mm}{!}{\includegraphics{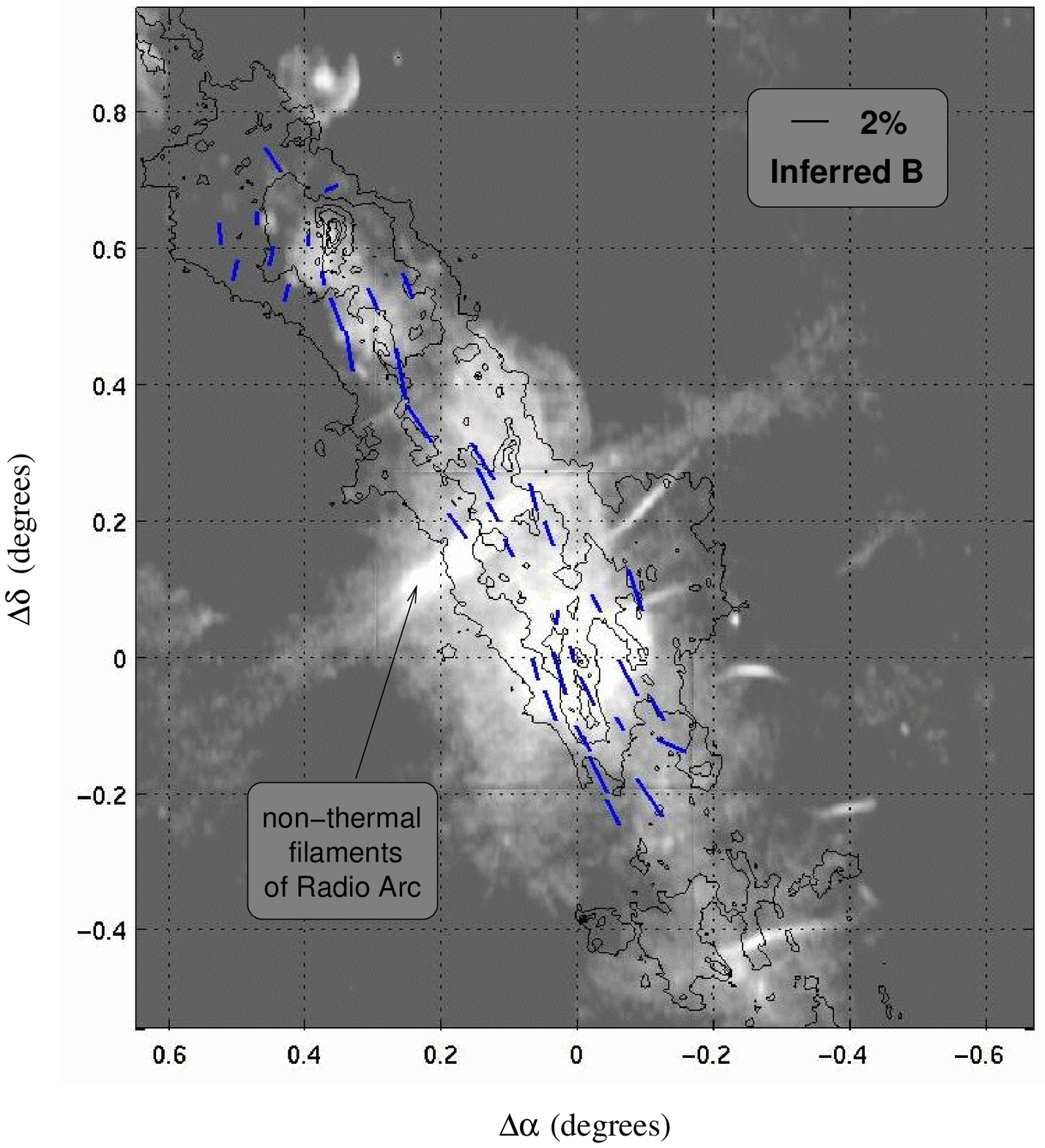}}
\caption{The diverse structure of radio emission near the Galactic
center at 3.6cm ({\it left}) from \cite{rg93} and 90cm ({\it right})
from \cite{lklh00}. Overlaid are the polarization observations at
sub-millimeter in \cite{agc+00, ncr+01}. The toroidal fields parallel
to the Galactic plane ({\it right}) are clearly detected in the
central molecular zone of 170~pc $\times$ 30~pc.
}
\end{figure}

As a matter of fact, recognition of the dipole fields in the
Galactic halo is fundamentally important. The magnetic fields of the dipole
in the Galactic center then must be very strong (see Fig.~3) and should show
many features. The filaments\cite{ymc84}, plumes \cite{dhr+98} and
even threads \cite{lme99} near the Galactic center all exactly reflect
these strong dipole fields. They are the locally illuminated flux tubes
in large-scale pervasive fields! Their locations, high linear
polarization of 50\% to 70\% \cite{lme99}, and their curvatures,
as well as their perpendicularity to the Galactic plane, all suggest
the dipole nature of the poloidal field structure. The strength of
the fields should of the order of mG \cite{ym87a}. 
As hinted by the direction of the local vertical fields, I predict
that the dipole fields in the Galactic center then should be directed
from the Northern pole to the Southern pole.

As well as the poloidal fields, Zeeman splitting \cite{yrg+96} and
polarization observations at sub-millimeter wavelength
\cite{ndd+00, ncr+01} have revealed strong
toroidal magnetic fields of 2 -- 4 mG in the central molecular zone
(see Fig.~4). Nova et al. found some evidence for the directions of
these toroidal fields \cite{ncr+01}, exactly the same field
configuration as we obtained for the halo field (see Fig.~3). This
has led us to believe that the A0 dynamo probably is working from
the GC to the halo. 

\section{Global structure of Galactic magnetic fields}
In the last decade, knowledge of the magnetic field structure of our
Galaxy has improved in many aspects through efforts in determining
pulsar RMs and mapping near the Galactic center. However, the story
is far from complete. Using the presently available information,
we can conclude that our Galaxy has such an odd symmetry of the
toroidal and poloidal fields in the Galactic halo and near the
Galactic center, showing that the dynamo is really working through 
the inductive effects of
fluid motions in the interstellar medium. However, bi-symmetric
spiral magnetic fields in the Galactic disk suggest that the disk
keeps some kind of memory of the field reversals from seed fields or 
the primordial fields. For any modeling to understand the origin
of cosmic rays, both the large
scale field in the disk and the halo should be considered.
Polarization observations of background radiation can directly
show the connections of large-scale disk fields to halo fields.

\begin{theacknowledgments}
I am very grateful to many colleagues, especially, Prof. R.N. Manchester
and Prof. G.J. Qiao for working together with me to improve the knowledge
on the magnetic fields of our Galaxy, some of which was presented here. 
Mr. Xu Dong is acknowledged for assistance on the field model simulations
and Dr. Tom Landecker for language editing.
I thank the partner group frame between Max-Planck Society
and Chinese Academy of Sciences for a long-term cooperation between
the MPIfR and NAOC.
My research in China is supported by the National Natural Science
Foundation of China and the National Key Basic Research Science
Foundation of China (NKBRSF G19990752).
\end{theacknowledgments}

\doingARLO[\bibliographystyle{aipproc}]
          {\ifthenelse{\equal{\AIPcitestyleselect}{num}}
             {\bibliographystyle{arlonum}}
             {\bibliographystyle{arlobib}}
          }
\bibliography{han}

\end{document}